\documentclass{iaus}
\usepackage{graphicx}
\usepackage{natbib}
\usepackage{aas_macros}
\newcommand{\msun}{\mbox{${\rm M}_{\odot}$ }}

\title[Theoretical DTDs] 
{Theoretical Delay Time Distributions}

\author[Gijs Nelemans, Silvia Toonen, Madelon Bours]   
{Gijs Nelemans$^1$
 \and Silvia Toonen$^1$ \and Madelon Bours$^1$}

\affiliation{$^1$Dept. Astrophysics/IMAPP, Radboud University Nijmegen, P.O.. Box 9010, 6500 GL Nijmegen, the Netherlands \\ email: {\tt nelemans@astro.ru.nl}}

\pubyear{2011}
\volume{281}  
\pagerange{x--xx}
\jname{Title of your IAU Symposium}
\editors{A.C. Editor, B.D. Editor \& C.E. Editor, eds.}
\begin{document}

\maketitle

\begin{abstract}
We briefly discuss the method of population synthesis to calculate
theoretical delay time distributions of type Ia supernova
progenitors. We also compare the results of the different research
groups and conclude that although one of the main differences in the
results for single degenerate progenitors is the retention efficiency
with which accreted hydrogen is added to the white dwarf core, this cannot explain all 
the differences.
\end{abstract}

\firstsection 
\section{Introduction}

Given the uncertainties in the theoretical derivation of what the
progenitors of type Ia supernovae (SNIa) are, statistical methods can
in principle be useful. Viable progenitor scenarios must not only
potentially produce a SNIa (i.e. an explosion that looks like the
observed SNIa), they must also occur often enough to explain the
observed/inferred SNIa rate. In order to do so, an estimate needs to
be made about the occurrence frequency of different types of
binaries. This can be done using population synthesis techniques
\citep[e.g.][]{2006LRR.....9....6P} as has been done by several groups
for the different SNIa progenitor scenarios
\citep[e.g.][]{yl98,2005ASSL..332..163Y,2004MNRAS.350.1301H,2006MNRAS.368.1893F,2009ApJ...699.2026R,2010A&A...515A..89M,2009MNRAS.395..847W,2010MNRAS.401.2729W}. In
the following we will briefly discuss some of the important
ingredients of population synthesis calculations, uncertainties and
possible mitigating efforts before showing a comparison of results of
different groups and an attempt to understand the differences.

\section{Population synthesis}

The basics concept of population synthesis is simple: for a (large)
sample of initial binaries (i.e. two Zero-Age Main Sequence stars
orbiting each other with a certain orbital period) the subsequent
evolution is determined. Folding those with assumed initial parameter
(masses, periods) distributions and if needed star formation
histories, any binary population can be synthesised.

Population synthesis calculations typically evolve so many binaries
that using a stellar evolution code that solves the structure
equations explicitly is not feasible. Even if that would be the case,
there are many binary evolution scenarios that cannot be computed with
such codes anyway. Therefore people use rapid fits to evolutionary
calculations or interpolations in pre-computed grids. This introduces
uncertainties, although these are typically small compared to other uncertainties.

More problematic is determining the outcome of binary interactions. If
one star in the binary fills its Roche lobe and starts to transfer
matter to the companion, one first has to determine if this will lead
to runaway mass transfer and the (likely?) start of a
``common-envelope'' phase. If not, the mass transfer is deemed
``stable'', but it still needs to be determined how much of the
transferred mass is accreted by the companion (likely depending on the
speed at which the mass is transferred) and how the remaining mass
leaves the system (i.e. how much angular momentum is lost along with
the mass). That will have a large impact on the further evolution of
the system.

In the context of SNIa, the two main progenitor scenarios are
carbon-oxygen white dwarfs that either merge with another
carbon-oxygen white dwarfs ("double degenerate" e.g. \citet{web84}) or
are accreting from a non-degenerate companion ("single degenerate"
e.g. \citet{1973ApJ...186.1007W}). The most important uncertainties
for population synthesis of these populations are the outcomes of mass
transfer and for the SD progenitors in addition the question which
white dwarf --- companion star configurations lead to sufficient
accretion to reach ignition conditions by approaching the
Chandrasekhar mass. These uncertainties are sufficiently large
that we simply assume or hope that the prescriptions we use about
these mass transfer phases in our population synthesis codes are
right. Instead we have to use available observational data to
calibrated the models.

\subsection{Comparison with observations}

\begin{figure}
\begin{center}
\includegraphics[width=9cm,angle=-90]{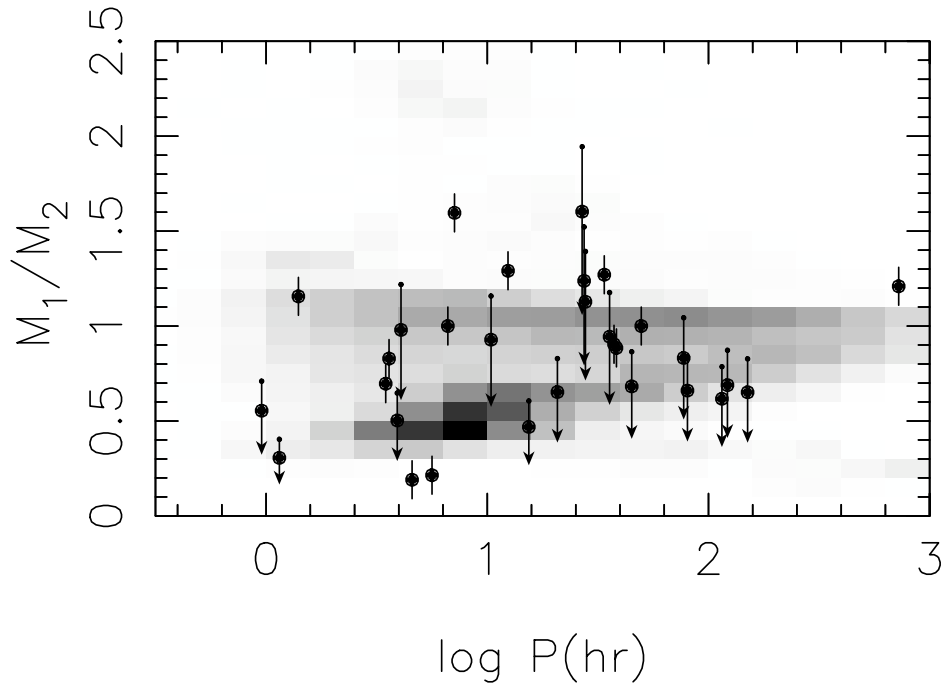}
 \caption{Period -- mass distribution of the observed double white
   dwarfs compared to the results of the population synthesis of
   Toonen et al. in prep. The simulation matches the observations
   reasonably well.}
   \label{figobs}
\end{center}
\end{figure}

Comparison of the simulated populations with the observed population
of (carbon-oxygen) white dwarfs with non-degenerate companions has not
be done yet to our knowledge. One of the reasons is that the 
observed population is severely biased by observational selection
effects that need careful modelling before the observations can be
compared to the population synthesis calculations. We are planning to
do that in a forthcoming paper. The population of close double white
dwarfs is much more homogeneous and we have compared the results of
our population synthesis calculations with the observations to
constrain the uncertain mass transfer phases \citep[e.g.][Toonen et
  al. in prep]{nyp+00}. In Fig.~\ref{figobs} we show this comparison
for our newest simulation of the double white dwarf population (Toonen
et al. in prep.).

There is a limit to the accuracy of this comparison when considering
DD progenitor scenarios, because non of the observed systems is likely
a double carbon-oxygen white dwarf with sufficient mass. However, the
agreement with the other double white dwarfs does give some confidence
in the simulation results.

\section{Delay time distributions of different groups}

\begin{figure}
  \includegraphics[angle=-90,width=0.9\columnwidth,clip]{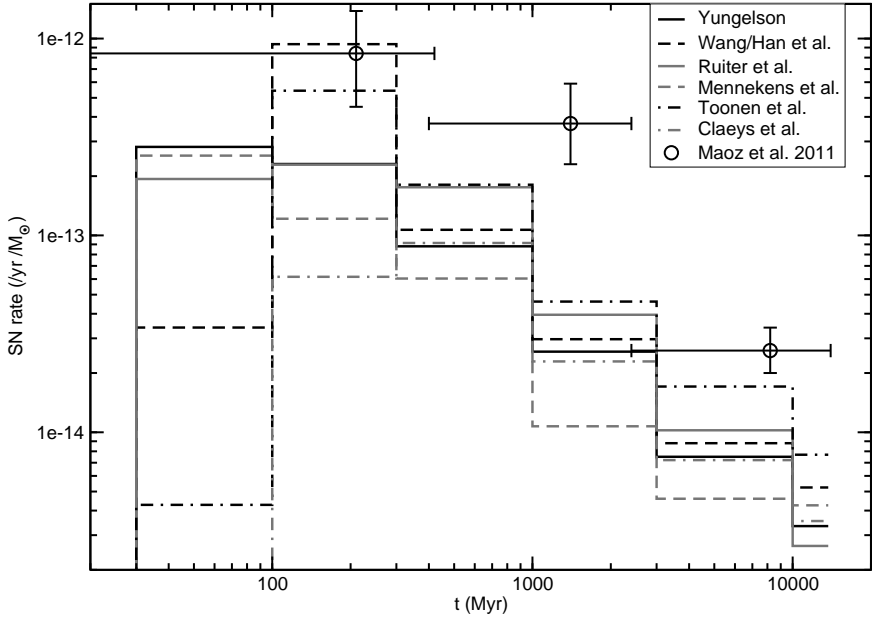}
  \caption{Rescaled DTDs for the DD scenario for the different groups.}
\label{fig:DTDs_DD}
\end{figure}

\begin{figure}
  \includegraphics[angle=-90,width=0.9\columnwidth,clip]{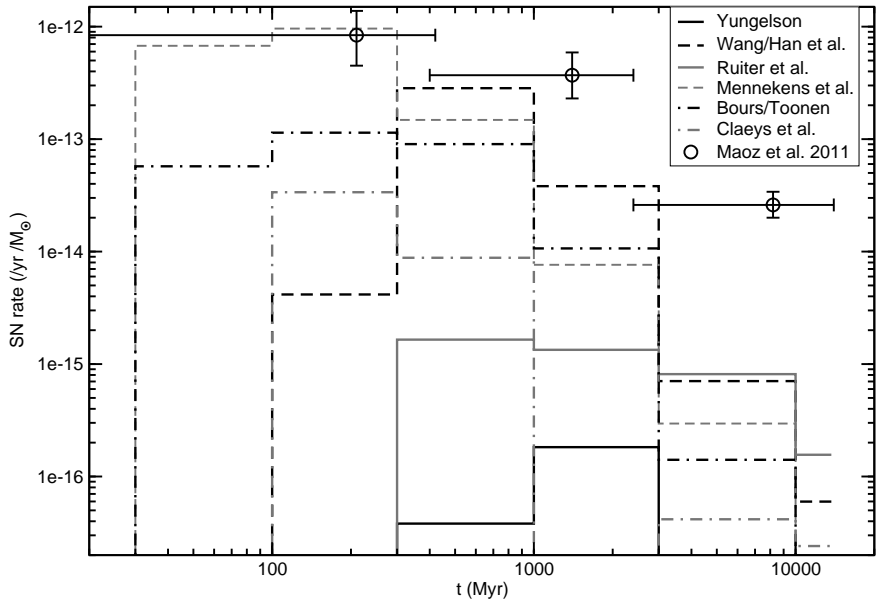}
  \caption{Rescaled DTDs for the SD scenario.}
\label{fig:DTDs_SD}
\end{figure}

Last year for the Lorentz center workshop "Observational signatures of
type Ia supernova progenitors", we asked a number of different
research groups to give us their data on the type Ia supernova
progenitors so that we could do a comparison of the results of the
different groups. All groups kindly accepted the invitation and were
very helpful in clarifying any queries regarding their methods and
normalisation. We decided to make the comparison of the so-called
"Delay Time Distribution" (DTD), which shows the SNIa rate per unit
mass as a function of time since a hypothetical instantaneous
starburst. For binary population synthesis calculations this is easy
to compute, the only issue hampering direct comparison between groups
is the normalisation. This depends on the assumed Initial Mass
Function, the percentage single stars and the initial distribution of
mass ratios and orbital periods. The groups involved are
\begin{itemize}
\item Lev Yungelson's work \citep[e.g.][]{2010AstL...36..780Y,2005ASSL..332..163Y,yl00}
\item The Yunnan group (Han, Wang, etc), see \citet{2010MNRAS.401.2729W,han08,2004MNRAS.350.1301H,han98}
\item The StarTrack code, Belczynski/Ruiter work
\citep[e.g.][]{2009ApJ...699.2026R,2008ApJS..174..223B}
\item The Brussels group \citep[see][]{2010A&A...515A..89M}
\item The Utrecht group, Claeys, Pols (e.g. Claeys, this proceedings)
\item Our SeBa code \citep[Toonen et al. (in prep),][]{nyp+00,pv96}
\end{itemize}

For the comparison we rescaled the results (if needed) to the
following assumptions: flat mass ratio distribution, a \citet{ktg93}
IMF, flat mass ratio distribution, flat distribution in $\log P_{\rm
  orb}$ and 50\% binaries. The conversion factors for the different
groups are given in table~\ref{tab:int_rates}.

In figures~\ref{fig:DTDs_DD} and \ref{fig:DTDs_SD} we show the
(binned) DTDs for the different groups, for the DD and SD scenarios
respectively. For the latter we only considered systems with hydrogen
rich donors, excluding helium rich donors
\citep[e.g.][]{2009MNRAS.395..847W}. We also plot the inferred data
points from \citet{2010arXiv1002.3056M}.  Most striking is that the
double degenerate channel shows roughly the same shape in all the
calculations, although with differences in the actual rates, but that
the single degenerate DTDs are all over the place.

Because it is rather hard for the eye to integrate the area in a log -
linear plot we show in table~\ref{tab:int_rates} the integrated rates
over a Hubble time (i.e. in total number of SNIa per \msun in a Hubble
time), together with the inferred rate from
\citet{2010arXiv1002.3056M}. It is clear that all population synthesis
calculations (with this normalisation!) produce too little SNIa.

\begin{table}
\caption{Normalisation factors and integrated SNIa rates in $10^{-4}
  \msun^{-1}$ for the different groups and the different channels. For
  comparison the integrated rate inferred from the observations by
  \citet{2010arXiv1002.3056M} is also given.}
\label{tab:int_rates}
\begin{center}
\begin{tabular}{lrrr}
\hline
Group & DD & SD & factor\\ \hline
Yungelson & 2.4 & 0.006 & 0.8 \\
Wang/Han & 4.4 & 2.8 & 0.77\\
Ruiter & 5.7 & 0.17 & 1?\\
Mennekens & 2.2 & 3.7& 0.62  \\
Claeys & 1.9 & 0.13 & 1\\
Toonen/Bours & 1.9 & 1.1 & 1 \\
\hline
\textbf{Observed} & \multicolumn{2}{c}{\textbf{23}} \\
\hline
\end{tabular}
\end{center}
\end{table}

\section{Can we understand the differences?}

\begin{figure}
  \includegraphics[width=0.9\columnwidth,clip]{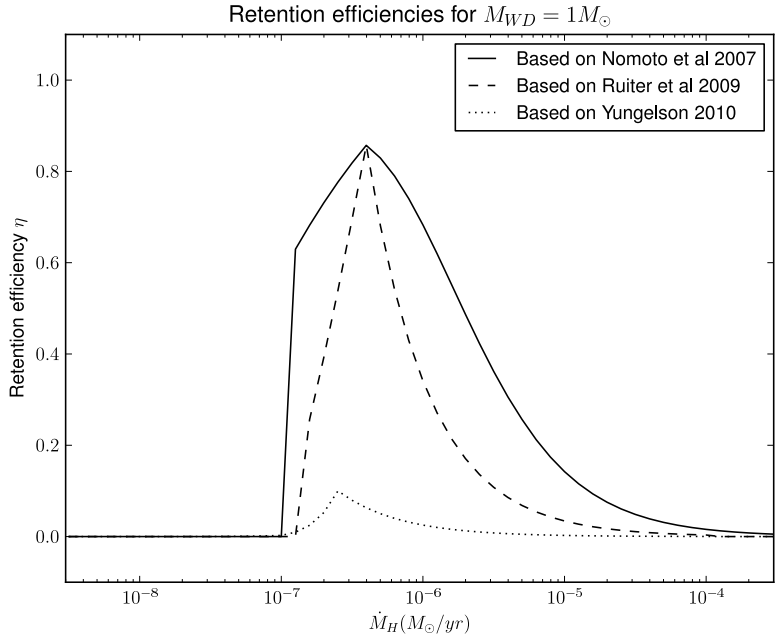}
  \caption{Retention efficiencies (i.e. efficiency with with accreted
    hydrogen is tuned into carbon and oxygen used by the different groups.}
\label{fig:retefs}
\end{figure}

\begin{figure}
  \includegraphics[angle=90,width=0.9\columnwidth,clip]{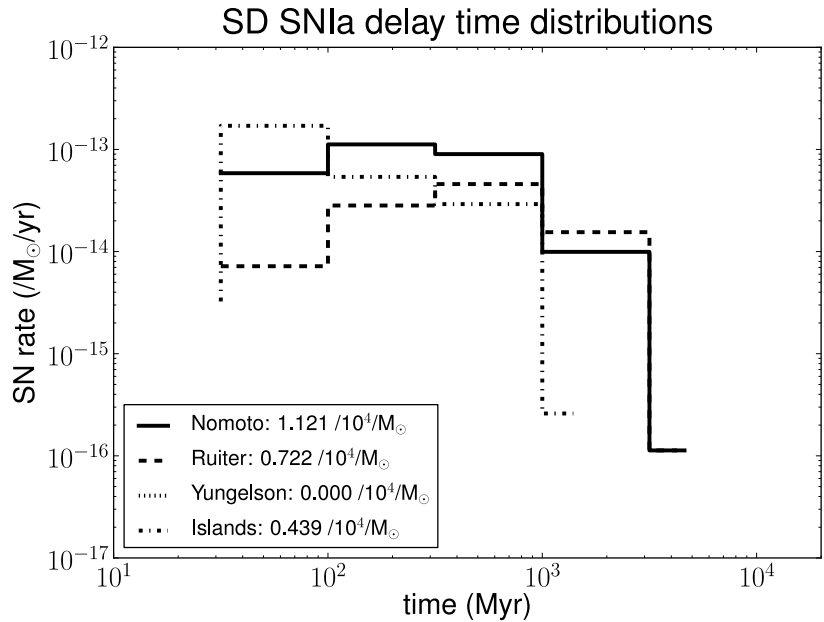}
  \caption{DTDs simulating the different approaches: ``islands'' for assuming the parameters that lead to SD SNIa to lie in the islands of \citet{2008ApJ...683L.127H}. The others are for the retention efficiencies shown in Fig.~\ref{fig:retefs} used by \citet{2010AstL...36..780Y,2005ASSL..332..163Y,yl00}, \citet{2009ApJ...699.2026R} and one that should reproduce the islands.}
\label{fig:DTDs_SD_sim}
\end{figure}

We investigated the most likely cause of the enormous discrepancy
between the groups for the SD scenario: the actual conditions in which
an accreting white dwarf can grow in mass and thus explode as a SNIa
(see Bours et al. in prep for more details). The different groups use
a different approach: some use so-called retention efficiencies (the
fraction of accreted hydrogen that in the end is added to the core of
the white dwarf as carbon and oxygen), some use progenitor ``islands''
in parameter space (of orbital period, white dwarf and companion
mass). We first calculated the SD DTD by using the islands as
calculated by \citet{2008ApJ...683L.127H}, the same approach as taken
by \citet{2010A&A...515A..89M}. We also implemented retention
efficiencies that are the basis of these islands, ones that are used
by \citet{2009ApJ...699.2026R} and ones that are used by
\citet{2010AstL...36..780Y} (see Fig.~\ref{fig:retefs}). The latter
two are based on calculations by
\citet{1995ApJ...445..789P,1996ApJS..105..145I}. The resulting DTDs of
these calculations are shown in Fig.~\ref{fig:DTDs_SD_sim}. As can be
seen, the different retention efficiencies do cause a large difference
in the integrated rates (with no systems left with the ``Yungelson''
efficiencies), as is the case in the comparison in the previous
section. However, the differences in the shape of the DTDs as found in
the previous section are not reproduces in our modelling. This suggest
that apart from the retention efficiencies other differences in the
codes, most likely in the treatment of the mass transfer phases
contribute significantly to the uncertainties in the theoretical DTDs.

\section{Conclusions}

We have shown that for the theoretical DTDs for the DD scenario there
are (limited) observational tests that can be made to constrain the
results. Interestingly, the DTDs of different research groups for the
DD scenario agree reasonably well. For the SD scenario the DTDs differ
wildly. This is partly due to the extreme uncertainty of the retention
efficiency of accreted hydrogen on carbon-oxygen white
dwarfs. However, a test in which this is varied while keeping the rest
of the population synthesis code fixed, shows that  this alone cannot
explain the differences in the SD DTDs between the different research
groups. Before the SD DTDs can be used to make any statements about
the likelihood of this scenario, the retention efficiencies have to be
determined better and the progenitor population has to be compared in
detail to the local observed sample.

\section{Acknowledgement}

It is a great pleasure to thank our colleagues Lev Yungelson, Bo Wang,
Zhanwen Han, Ashley Ruiter, Nicky Mennekens and Joke Claeys for
sharing their data with us and for numerous stimulating discussions.

%
\bibliographystyle{nato}  
\bibliography{journals,binaries}                

\begin{thebibliography}{}

\bibitem[\protect\astroncite{{Belczynski} et~al.}{2008}]{2008ApJS..174..223B}
{Belczynski}, K., {Kalogera}, V., {Rasio}, F.~A., {Taam}, R.~E., {Zezas}, A.,
  {Bulik}, T., {Maccarone}, T.~J., and {Ivanova}, N. 2008,
 {\em \apjs} {\bf 174}, 223

\bibitem[\protect\astroncite{{F{\"o}rster} et~al.}{2006}]{2006MNRAS.368.1893F}
{F{\"o}rster}, F., {Wolf}, C., {Podsiadlowski}, P., and {Han}, Z. 2006,
 {\em \mnras} {\bf 368}, 1893

\bibitem[\protect\astroncite{{Hachisu} et~al.}{2008}]{2008ApJ...683L.127H}
{Hachisu}, I., {Kato}, M., and {Nomoto}, K. 2008,
 {\em \apjl} {\bf 683}, L127

\bibitem[\protect\astroncite{Han}{1998}]{han98}
Han, Z. 1998,
 {\em MNRAS} {\bf 296}, 1019

\bibitem[\protect\astroncite{{Han}}{2008}]{han08}
{Han}, Z. 2008,
 {\em \apjl} {\bf 677}, L109

\bibitem[\protect\astroncite{{Han} and
  {Podsiadlowski}}{2004}]{2004MNRAS.350.1301H}
{Han}, Z. and {Podsiadlowski}, P. 2004,
 {\em \mnras} {\bf 350}, 1301

\bibitem[\protect\astroncite{{Iben} and {Tutukov}}{1996}]{1996ApJS..105..145I}
{Iben}, Jr., I. and {Tutukov}, A.~V. 1996,
 {\em \apjs} {\bf 105}, 145

\bibitem[\protect\astroncite{{Kroupa} et~al.}{1993}]{ktg93}
{Kroupa}, P., {Tout}, C.~A., and {Gilmore}, G. 1993,
 {\em \mnras} {\bf 262}, 545

\bibitem[\protect\astroncite{{Maoz} et~al.}{2010}]{2010arXiv1002.3056M}
{Maoz}, D., {Mannucci}, F., {Li}, W., {Filippenko}, A.~V., {Della Valle}, M.,
  and {Panagia}, N. 2010,
 {\em \mnras} {\bf {412}}, 1508

\bibitem[\protect\astroncite{{Mennekens} et~al.}{2010}]{2010A&A...515A..89M}
{Mennekens}, N., {Vanbeveren}, D., {De Greve}, J.~P., and {De Donder}, E. 2010,
 {\em \aap} {\bf 515}, 89

\bibitem[\protect\astroncite{Nelemans et~al.}{2001}]{nyp+00}
Nelemans, G., Yungelson, L.~R., Portegies~Zwart, S.~F., and Verbunt, F. 2001,
 {\em A\&A} {\bf 365}, 491

\bibitem[\protect\astroncite{Portegies~Zwart and Verbunt}{1996}]{pv96}
Portegies~Zwart, S.~F. and Verbunt, F. 1996,
 {\em A\&A} {\bf 309}, 179

\bibitem[\protect\astroncite{{Postnov} and
  {Yungelson}}{2006}]{2006LRR.....9....6P}
{Postnov}, K.~A. and {Yungelson}, L.~R. 2006,
 {\em Living Reviews in Relativity} {\bf 9}, 6

\bibitem[\protect\astroncite{{Prialnik} and
  {Kovetz}}{1995}]{1995ApJ...445..789P}
{Prialnik}, D. and {Kovetz}, A. 1995,
 {\em \apj} {\bf 445}, 789

\bibitem[\protect\astroncite{{Ruiter} et~al.}{2009}]{2009ApJ...699.2026R}
{Ruiter}, A.~J., {Belczynski}, K., and {Fryer}, C. 2009,
 {\em \apj} {\bf 699}, 2026

\bibitem[\protect\astroncite{{Wang} et~al.}{2010}]{2010MNRAS.401.2729W}
{Wang}, B., {Li}, X., and {Han}, Z. 2010,
 {\em \mnras} {\bf 401}, 2729

\bibitem[\protect\astroncite{{Wang} et~al.}{2009}]{2009MNRAS.395..847W}
{Wang}, B., {Meng}, X., {Chen}, X., and {Han}, Z. 2009,
 {\em \mnras} {\bf 395}, 847

\bibitem[\protect\astroncite{Webbink}{1984}]{web84}
Webbink, R.~F. 1984,
 {\em ApJ} {\bf 277}, 355

\bibitem[\protect\astroncite{{Whelan} and {Iben}}{1973}]{1973ApJ...186.1007W}
{Whelan}, J. and {Iben}, I.~J. 1973,
 {\em \apj} {\bf 186}, 1007

\bibitem[\protect\astroncite{{Yungelson}}{2005}]{2005ASSL..332..163Y}
{Yungelson}, L.~R. 2005,
 in E.~M. {Sion}, S. {Vennes}, and H.~L. {Shipman} (eds.), {\em White dwarfs:
  cosmological and galactic probes}, Vol. 332 of {\em Astrophysics and Space
  Science Library}, Astrophysics and Space Science Library, pp 163--173

\bibitem[\protect\astroncite{{Yungelson}}{2010}]{2010AstL...36..780Y}
{Yungelson}, L.~R. 2010,
 {\em Astronomy Letters} {\bf 36}, 780

\bibitem[\protect\astroncite{Yungelson and Livio}{1998}]{yl98}
Yungelson, L.~R. and Livio, M. 1998,
 {\em ApJ} {\bf 497}, 168

\bibitem[\protect\astroncite{Yungelson and Livio}{2000}]{yl00}
Yungelson, L.~R. and Livio, M. 2000,
 {\em ApJ} {\bf 528}, 108

\end{thebibliography}

\end{document}